\documentclass[12pt, prd, showpacs]{revtex4}
\usepackage{amssymb}
\usepackage{amsmath}

\input{tcilatex}

\begin{document}

\title{Acceleration of particles by acceleration horizons}
\author{O. B. Zaslavskii}
\affiliation{Department of Physics and Technology, Kharkov V.N. Karazin National
University, 4 Svoboda Square, Kharkov 61022, Ukraine}
\email{zaslav@ukr.net }

\begin{abstract}
We consider collision of two particles in the vicinity of the extremal
acceleration horizon (charged or rotating) that includes the
Bertotti-Robinson space-time and the geometry of the Kerr throat. It is
shown that the energy in the centre of mass frame $E_{c.m.}$ can become
indefinitely large if parameters of one of the particles are fine-tuned, so
the Ba\~{n}ados-Silk-West (BSW) effect manifests itself. There exists
coordinate transformation which brings the metric into the form free of the
horizon. This leads to some paradox since (i) the BSW effect exists due to
the horizon, (ii) $E_{c.m.}$ is a\ scalar and cannot depend on the frame.
Careful comparison of near-horizon trajectories in both frames enables us to
resolve this paradox. Although globally the space-time structure of the
metrics with acceleration horizons and black holes are completely different,
locally the vicinity of the extremal black hole horizon can be approximated
by the metric of the acceleration one. The energy of one particle from the
viewpoint of the Kruskal observer (or the one obtained from it by finite
local boost) diverges although in the stationary frame energies of both
colliding particles are finite. This suggests a new explanation of the BSW
effect for black holes given from the viewpoint of an observer who crosses
the horizon. It is complementary to the previously found explanation from
the point of view of a static or stationary observer.
\end{abstract}

\keywords{BSW effect, acceleration horizon, Bertotti-Robinson space-time}
\pacs{04.70.Bw, 97.60.Lf }
\maketitle

\section{Introduction}

The Ba\~{n}ados-Silk-West effect (denoted hereafter the BSW effect),
discovered in 2009 \cite{ban}, still attracts much attention. It consists in
the possibility of getting indefinitely large energy $E_{c.m.\text{ }}$ in
the centre of mass frame of two particles colliding near the black hole
horizon. As $E_{c.m.\text{ }}$can be made as large as one likes, this leads
to the possibility of creation of high-energetic and/or massive particles
and opening new channels of reaction forbidden in laboratory conditions.

The basic features of the BSW effect can be summarized as follows: (i)
collision occurs near the horizon, so the presence of the horizon is
essential, (ii) the Killing energy $E_{i}$ ($i=1,2$) of each particle is
finite but $E_{c.m.\text{ }}$ is as large as one likes, (iii) one of
particles has the fine-tuned relationship between the energy $E_{1}$ and its
angular momentum or electric charge (so-called critical particle), whereas
the second particle is "usual" in the sense that its parameters are
arbitrary (not fine-tuned). A simple kinematic explanation of the BSW effect
was suggested in \cite{k}. It was done from the point of view of the
stationary (or static) observer who resides outside the horizon. Meanwhile,
an alternative explanation should exist from the viewpoint of the observer
who crosses the horizon \ Obviously, both observers should agree that $%
E_{c.m.\text{ }}$ grows unbound since it is a scalar and its value cannot
depend on frame. However, more careful inspection reveals some paradox here.
Indeed, such an observer (say, the Kruskal one) does not see anything
special (unless he carries out some very subtle geometric measurements) when
he crosses the horizon. Therefore, it seems that aforementioned basic point
(i) fails. How can one explain the effect of unbound $E_{c.m.\text{ }}$in
this frame?

The situation becomes more pronounced if, instead of a black hole, one
considers a so-called acceleration horizon. It arises due to a pure
kinematic effect and can be removed by passing into a frame connected with a
different observer. One of the most known examples is the Rindler metric. If
a suitable coordinate transformation is performed, the standard metric of
the Minkowskian space-time is revealed that, obviously, does not have a
horizon. Another example, more relevant in the context of the BSW effect is
the Bertotti-Robinson space-time \cite{ber}, \cite{rob}. If the metric of an
acceleration horizon is such that the lapse function vanishes on some
surface, the BSW effect should take place there. And, this makes the
aforementioned paradox even more pronounced for acceleration horizons since
the horizon is present in one frame (with points (i) - (iii) satisfied) and
is absent in the other one. The vicinity of a true black hole horizon can be
approximately described by the metric of an acceleration one. Therefore, the
latter type of horizons is a very useful tool for better understanding the
kinematics of the BSW effect. Thus, in what follows we should distinguish
(i) two kinds of horizons and, within each kind, (ii) two different frames.

Meanwhile, new questions arise here. The acceleration horizon can be
eliminated, whereas the same is not true for the physical horizon of a black
hole - say, the Kerr one, where the BSW effect was first discovered \cite%
{ban}. Therefore, on the face it, one could naively expect the crucial
difference between this effect near black holes and acceleration horizons.
We show that this is not so. It is the local properties of the metric (that
are similar for acceleration and black hole horizons) but not the global
character of causal structure of space-time (that are different for both
types of horizons) which are relevent in the given context. This ensures
continuity between both kinds of the BSW effect that exists due to the fact
that a black hole metric can be approximated by the metric of an
acceleration horizon.

Comparison of both frames and properties of trajectories of particles helps
us to understand better the kinematics of the BSW effect. In \cite{k},
kinematic explanation was done from the viewpoint of static (stationary)
observer who orbits the horizon but does not cross it. Now, another
explanation is suggested from the viewpoint of an observer who crosses the
horizon. It applies to both black hole and acceleration horizons.

We begin with the Bertotti-Robinson space-time since it looks simpler than
its rotating counterpart. However, as they have much in common, the obtained
results are extended to the rotating BR in a straightforward manner. (It is
worth noting that collision in the background of rotating acceleration
horizons were considered recently in \cite{g} but our conclusions are
qualitatively different - see below for more details.)

Up to now, we considered acceleration horizons as a useful tool for
description of black holes. Meanwhile, such a kind of horizon is of interest
on its own right. For instance, the Bertotti-Robinson space-time and its
rotational analogue appear in the processes of different limiting
transitions in the context of gravitational thermal ensembles \cite{hm} and
can be relevant in the context of the AdS/CFT correspondence \cite{mald}, 
\cite{b} or the Kerr/CFT one \cite{kcft}. They are also encountered in many
other physical contexts connected with nonlinear electrodynamics \cite{mat},
conformal mechanics \cite{conf}, limiting transitions from rapidly rotating
discs to black holes \cite{disc}, etc. Acceleration horizons approximately
describe an infinite throat of the extremal Kerr, Reissner-Nordstr\"{o}m or
Kerr-Newman black holes. Therefore, information about properties of motion
in such background can be useful even in astrophysical context. Falling
matter can spin up a black hole significantly \cite{bn}, \cite{ln}, so that
it can acquire an angular momentum close to its mass and become almost
extremal \cite{th} - \cite{frolovkerr}.

Some reservations are in order. We do not consider here the force of
gravitational radiation which seemed to restrict the BSW effect \cite{berti}%
, \cite{ted}. This is not only because such neglect is reasonable in the
main approximation \cite{insp} but also since under rather general
assumptions about the force, the BSW effect survives in principle \cite{rad}%
. There are also astrophysical restrictions \cite{mc} but they depend
crucially on the type of the physical situation (see more on this in \cite%
{com}). Anyway, the BSW effect is an interesting phenomenon on its own, and
its subtleties deserve careful studies.

Throughout the paper we use units in which fundamental constants are $G=c=1$%
. The paper is organized as follows. In Sec. II, we give basic formulas for
different forms of the Bertotti-Robinson metric and its relation to the
Reissner-Nordstr\"{o}m one. In Sec. III, we briefly outline the BSW effect
for radial motion of particles in spherically symmetric space-times. In Sec.
IV, we formulate the paradox of two frames and suggest its resolution.
General discussion of properties of motion responsible for the explanation
of this paradox is given in Sec. V. To reveal the essence of matter, in Sec.
VI we exploit a very simple model of particles moving in the flat
space-time. To compare the BSW effect near black holes and near acceleration
horizons, in Sec. VII we apply the Kruskal transformations to metrics with
both types of horizon. In Sec. VIII - X we consider general properties of
the acceleration horizons for axially symmetric rotating space-times. In
Sec. XI, we give basic results for the BSW effect in such space-times. In
Sec. XII, we apply this approach to collisions near the throat obtained from
the Kerr solution. In Sec. XIII, we critically review the previous attempt
of consideration of collisions in the throat geometry and explain why the
BSW effect was overlooked there. Sec. XIV is devoted to general discussion
of the results. In Appendix, we list useful formulas for the transformation
of the electric potential between two different frames in the
Bertotti-Robinson space-time.

\section{Bertotti-Robinson space-time}

Let us consider the spherically symmetric metric of the form 
\begin{equation}
ds^{2}=-fdt^{2}+\frac{dr^{2}}{f}+r^{2}d\Omega ^{2}\text{, }d\Omega
^{2}=d\theta ^{2}+\sin ^{2}\theta d\phi ^{2}\text{.}  \label{mf}
\end{equation}

The equations of motion of a test particle having the mass $m$ and the
charge $q$ read%
\begin{equation}
m\dot{t}=\frac{X}{f}\text{,}  \label{t}
\end{equation}%
\begin{equation}
m\dot{r}=\pm Z\text{, }Z=\sqrt{X^{2}-m^{2}f}\text{,}  \label{r}
\end{equation}%
\begin{equation}
X=E-q\varphi \text{,}  \label{X}
\end{equation}%
$\varphi $ is the electric potential, dot denotes differentiation with
respect to the proper time $\tau $. The quantity $X$ can be also written as $%
X=mu_{0}$, where $u^{\mu }$ is the four-velocity.

If $f=(1-\frac{r_{+}}{r})^{2}$, the metric describes the extremal
Reissner-Nordstr\"{o}m (RN) black hole with the horizon at $r=r_{+}$. Its
electric charge is $Q=r_{+}$, the electric potential%
\begin{equation}
\varphi =\frac{r_{+}}{r}.  \label{g1}
\end{equation}%
One can make the substitution%
\begin{equation}
r=r_{+}+\lambda x\text{, }t=\frac{\tilde{t}}{\lambda }  \label{rx}
\end{equation}%
and take the limit $\lambda \rightarrow 0$. Then, we obtain the metric%
\begin{equation}
ds^{2}=-dt^{2}\frac{x^{2}}{r_{+}^{2}}+r_{+}^{2}\frac{dx^{2}}{x^{2}}%
+r_{+}^{2}d\Omega ^{2}\text{,}  \label{br}
\end{equation}%
where, for simplicity, we omitted tilde in the notation of the time variable.

As $\varphi $ is the time component of the four-potential, the new potential
is $\varphi _{new}=\frac{\varphi _{old}}{\lambda }$, where $\varphi _{old}$
is given by (\ref{g1}). To make the limiting transition $\lambda \rightarrow
0$ well-defined, we can change the arbitrary constant in the definition of
the potential in such a way that, say, $\varphi _{new}=1$ at the horizon.
Thus we can write%
\begin{equation}
\varphi _{new}=\frac{r_{+}-r}{\lambda r}+1\text{.}
\end{equation}

Then, one can perform the transition in question and obtain that in the
Taylor expansion of $\varphi _{new}$ with respect to $\lambda x$ (\ref{rx})
all terms of the order $x^{2}$ and higher vanish, so the electric potential
is%
\begin{equation}
\varphi =1-x,  \label{phi}
\end{equation}%
where we omitted subscript "new" for simplicity.

This is the Bertotti-Robinson (BR) space-time \cite{ber}, \cite{rob} which
is the exact solution of the Einstein-Maxwell equations. Alternatively, one
can simply use the approximate expansions near the horizon%
\begin{equation}
f\approx x^{2}\text{, }\varphi \approx 1-x\text{, }x=\frac{r-r_{+}}{r_{+}},
\end{equation}%
truncate them and obtain the metric (\ref{br}) and the potential (\ref{phi})
It is invariant with respect to scaling $x\rightarrow \lambda x$, $%
t\rightarrow \frac{t}{\lambda }$. Hereafter, we take for simplicity $r_{+}=1$%
.

Thus the BR\ metric can be considered either as an approximation to the RN
one in the near-horizon region or by itself. The metric (\ref{br}) possesses
the horizon at $x=0$. However, in contrast to the RN metric, this is not a
black hole horizon but is a so-called acceleration horizon. It appears due
to the choice of frame and can be removed globally in the corresponding one,
so this is a pure kinematic effect. Indeed, let us make the substitution

\begin{equation}
x=\sqrt{1+y^{2}}\cos \tilde{t}+y,  \label{xty}
\end{equation}%
\begin{equation}
t=\frac{\sqrt{1+y^{2}}\sin \tilde{t}}{\sqrt{1+y^{2}}\cos \tilde{t}+y}.
\label{tty}
\end{equation}%
This transformation is similar to that for the rotational counterpart of the
BR space-time \cite{b}. The inverse transformations reads%
\begin{equation}
y=\frac{1}{2}(x+xt^{2}-\frac{1}{x}),  \label{yx}
\end{equation}%
\begin{equation}
\sin \tilde{t}=\frac{xt}{\sqrt{1+\frac{1}{4}(xt^{2}+x-\frac{1}{x})^{2}}},
\end{equation}%
\begin{equation}
\cos \tilde{t}=\frac{1}{2}\frac{(x+\frac{1}{x}-xt^{2})}{\sqrt{1+\frac{1}{4}%
(xt^{2}+x-\frac{1}{x})^{2}}}.
\end{equation}

Then, the metric takes the form%
\begin{equation}
ds^{2}=-d\tilde{t}^{2}(1+y^{2})+\frac{dy^{2}}{1+y^{2}}+d\Omega ^{2}.
\label{y}
\end{equation}

Regarding the electromagnetic potential $A_{\mu }$, the situation is less
straightforward. Calculations are direct but rather cumbersome. As a result,
it turns out that if we start with the gauge (\ref{g1}), all components of $%
A_{\mu }$ are nonzero, including spatial ones $A_{i}$ and, moreover, the
electromagnetic potential depends on time. The final expressions are rather
lengthy. However, one can perform transformation to a new gauge in which $%
A_{i}=0$ and the new potential is 
\begin{equation}
\tilde{\varphi}=1-y  \label{fy}
\end{equation}%
(see Appendix for details). Instead of that, one can start from (\ref{y})
directly and check that with the potential (\ref{phi}), the Maxwell
equations $F_{;\nu }^{\mu \nu }=0$ are satisfied (here $F_{\mu \nu
}=\partial _{\mu }A_{\nu }-\partial _{\nu }A_{\mu }$ is the electromagnetic
field tensor), semicolon denotes covariant derivative.

The BR space-time possesses two inequivalent Killing vectors $\xi ^{\mu }$
and $\tilde{\xi}^{\mu }$ corresponding to time translations in $t$ and $%
\tilde{t}$. The vector $\xi ^{\mu }=(1,0,0,0)$ in the frame (\ref{br}) and $%
\tilde{\xi}^{\mu }=(1,0,0,0)$ in the frame (\ref{y}). As both frames are
different, $\xi ^{\mu }\neq $ $\tilde{\xi}^{\mu }$. Corresponding energies
are $E=-P_{\mu }\xi ^{\mu }$ and $\tilde{E}=-P_{\mu }\tilde{\xi}^{\mu }$,
where $P_{\mu }=mu_{\mu }+eA_{\mu }$ is the generalized momentum. The
energies also may be different in general. (Actually, there is one more
time-like Killing vector in the BR space-time \cite{lap} but it is
irrelevant in our context.)

We restrict ourselves to pure radial motion. Then, the equations of motion
give us%
\begin{equation}
m\frac{d\tilde{t}}{d\tau }=\frac{\tilde{X}}{1+y^{2}}\text{,}  \label{tt}
\end{equation}%
\begin{equation}
m\frac{dy}{d\tau }=\tilde{Z}\text{,}  \label{yt}
\end{equation}%
\begin{equation}
\tilde{X}=\tilde{E}-q+qy,  \label{xy}
\end{equation}%
\begin{equation}
\tilde{Z}=\sqrt{\tilde{X}^{2}-m^{2}(1+y^{2})}.  \label{zy}
\end{equation}

\section{BSW effect: general formulas}

For one particle, the standard textbook formula states that the $%
E^{2}=-p^{\mu }p_{\mu }$ where~$p^{\mu }=mu^{\mu }$ is the four-momentum, $E$
is the energy, $u^{\mu }=\frac{dx^{\mu }}{d\tau }$ is the four-velocity. For
two colliding particles, in the point of collision one can define the energy
in the centre of mass frame as%
\begin{equation}
E_{c.m.}^{2}=-\left( m_{1}u_{1}^{\mu }+m_{2}u_{2}^{\mu }\right) \left(
m_{1}u_{1\mu }+m_{2}u_{2\mu }\right) .\text{ }
\end{equation}%
Then,%
\begin{equation}
E_{c.m.}^{2}=m_{1}^{2}+m_{2}^{2}+2m_{1}m_{2}\gamma
\end{equation}%
where the Lorentz factor of relative motion%
\begin{equation}
\gamma =-u_{1}^{\mu }u_{2\mu }\text{,}  \label{gamma}
\end{equation}%
$\left( u^{\mu }\right) _{i}$ is the four-velocity of the i-th particle
(i=1,2).

Applying eqs. (\ref{t}), (\ref{r}), one can obtain%
\begin{equation}
\gamma =\frac{X_{1}X_{2}-Z_{1}Z_{2}}{m_{1}m_{2}f}\text{.}  \label{xz}
\end{equation}

The BSW effect happens when collision occurs near $r_{+}$, where $%
f\rightarrow 0$. It requires that for one particle (say, particle 1) the
relation $X_{1}(r_{+})=0$ to hold (so-called "critical" particle), whereas
for particle 2 (so-called "usual" one) $X_{2}(r_{+})\neq 0$. As a result,%
\begin{equation}
\gamma \approx \frac{X_{2}(r_{+})}{m_{1}m_{2}\sqrt{f}}(E_{1}-\sqrt{%
E_{1}^{2}-m_{1}^{2}})
\end{equation}%
becomes indefinitely large when the point of collision $r\rightarrow r_{+}$
(see \cite{jl} for details). In the point of collision $x_{c},$ the metric
function corresponding to (\ref{br}) behaves according to $f\sim x_{c}^{2}$
so%
\begin{equation}
\gamma \sim \frac{1}{x_{c}}\text{.}  \label{gac}
\end{equation}

\section{The paradox of two frames}

Meanwhile, for the space-time under discussion we are faced with the
following paradox. On one hand, in the form (\ref{br}), there is the horizon
where $f=x^{2}\rightarrow 0$, so the general scheme predicts the BSW effect.
From the other hand, it is seen from (\ref{y}) that there is no horizon, so
it seems obvious that there is no reason to anticipate this effect. It
follows from (\ref{tt}), (\ref{yt}) and (\ref{xz}) that%
\begin{equation}
\gamma =\frac{\tilde{X}_{1}\tilde{X}_{2}-\tilde{Z}_{1}\tilde{Z}_{2}}{1+y^{2}}%
\text{,}  \label{gay}
\end{equation}%
where $\tilde{X}$ and $\tilde{Z}$ are given by (\ref{xy}) and (\ref{zy})$.$
Hereafter, we assume that $m_{1}=m_{2}=1$. For any fixed energies $\tilde{E}%
_{1,2}$ and for any $y$ the Lorentz factor $\gamma $ is bounded. However, as
it is a scalar, it is impossible to have $\gamma $ unbounded in one frame
and perfectly bounded in another one.

To gain insight into this issue, we will relate the characteristics of
particles in both frames (critical particle 1 and usual particle 2). This
will be done for two kinds of particles separately. In what follows, we use
the terms "critical" and "usual" particles with respect to their properties
in the frame (\ref{br}). We also use the term "horizon" for the surface that
in frame (\ref{y}) corresponds to $x=0$ in frame (\ref{br}).

\subsection{Critical particle}

By definition, it means $X_{1}(r_{+})=0$. Then, it follows from (\ref{X})
and (\ref{phi}) that $E_{1}=q_{1}\equiv q$, 
\begin{equation}
X_{1}=E_{1}x=qx.  \label{cx}
\end{equation}

Equations of motion (\ref{t}), (\ref{r}) give us (hereafter $m_{1}=1$)%
\begin{equation}
x_{1}=x_{0}\exp (-\lambda \tau )\text{, }\lambda =\sqrt{q^{2}-1}\text{,}
\label{crit}
\end{equation}%
\begin{equation}
t_{1}=\frac{q\exp (\lambda \tau )}{x_{0}\lambda }=\frac{q}{\lambda x_{1}}%
\text{,}  \label{tc}
\end{equation}%
the constant of integration in (\ref{tc}) is set to zero; it is implied that 
$q>1$. Direct calculations based on coordinate transformations (\ref{xty}), (%
\ref{tty}) show that for the critical particle%
\begin{equation}
\tilde{X}_{1}=E_{1}y_{1}\text{, }\tilde{E}_{1}=E_{1}=q\text{.}  \label{xycr}
\end{equation}%
\begin{equation}
y_{1}=\frac{1}{2}(\frac{1}{x_{1}}\frac{1}{\lambda ^{2}}+x_{1})\text{.}
\label{y1}
\end{equation}

\subsection{Usual particle}

In what follows, we will restrict ourselves to the case when particle 2 is
neutral since it makes no qualitative difference but simplifies formulas
significantly. Then, in frame (\ref{br}) equations of motion have the
solution%
\begin{equation}
x_{2}=-E_{2}\sin \tau \text{,}  \label{xuE}
\end{equation}%
\begin{equation}
t_{2}=-\frac{1}{E_{2}}\cot \tau +t_{0}=\frac{1}{x}\sqrt{1-\frac{x^{2}}{%
E_{2}^{2}}}+t_{0}\text{.}  \label{tu}
\end{equation}

Here, we assume that $\tau =0$ at the moment of crossing the horizon, $\tau
<0$ before that. Then, using (\ref{yx}) one can see that in frame (\ref{y})%
\begin{equation}
y_{2}=t_{0}\cos \tau -p\sin \tau =\frac{p}{E_{2}}x+t_{0}\sqrt{1-\frac{x^{2}}{%
E_{2}^{2}}}\text{,}  \label{py}
\end{equation}%
\begin{equation}
p=\frac{1}{2}[(E_{2}-\frac{1}{E_{2}})+E_{2}t_{0}^{2}]\text{.}  \label{p}
\end{equation}

Correspondingly, the quantity $\tilde{X}\,_{2}\ $that appears in equation of
motion (\ref{tt}), (\ref{yt}) is equal to%
\begin{equation}
\tilde{X}_{2}=\tilde{E}_{2}=\frac{1}{2}(E_{2}+\frac{1}{E_{2}}+t_{0}^{2}E_{2})%
\text{.}  \label{Y}
\end{equation}

Thus we see that transformation of the quantity $X$ looks very different for
the critical and usual particles. For the critical particle $X_{1}$
coincides in both frames, but for a usual one this is not so.

\subsection{Explanation of the paradox}

At the first glance, nothing compels us to notice the BSW effect now since
for fixed $E$, $t_{0}$ and any $y$ described by eq. (\ref{py}) the Lorentz
factor $\gamma $ (\ref{gay}) is finite. However, the nontrivial point
consists in the proper account of the fact that (i) both particles should
meet in the same point and (ii) this point should be near the horizon. We
will see that combination of (i) and (ii) results in large constant of
integration $t_{0}$ and the very large energy $\tilde{E}_{2}\sim t_{0}^{2}$
which grows even faster that is crucial for calculation of $\gamma $.

Indeed, the event of collision implies that in the corresponding point
coordinates of two particles coincide in both frames:%
\begin{equation}
t_{1}=t_{2}\text{, }x_{1}=x_{2}\text{, }  \label{c1}
\end{equation}%
\begin{equation}
\tilde{t}_{1}=\tilde{t}_{2}\text{, }y_{1}=y_{2}.  \label{c2}
\end{equation}

Let collision occur at some small value of $x=x_{c}$. Then, it follows from (%
\ref{tc}), (\ref{tu}) that%
\begin{equation}
t_{0}=\frac{E-\lambda }{x_{c}\lambda }+O(x_{c})  \label{t0}
\end{equation}

Thus, for $x_{c}\rightarrow 0$ we also have%
\begin{equation}
t_{0}\sim x_{c}^{-1}\rightarrow \infty \text{.}  \label{tx}
\end{equation}

It also follows from (\ref{p}), (\ref{Y}) that%
\begin{equation}
p\sim \tilde{X}_{2}\sim t_{0}^{2}\sim \frac{1}{x_{c}^{2}}\rightarrow \infty
\label{p1}
\end{equation}%
where we took into account that it is particle 2 which is usual.

Near the point of collision $y=y_{c}$ we have from (\ref{y1}) and (\ref{py})
that%
\begin{equation}
y_{c}\sim \frac{1}{x_{c}}\rightarrow \infty \text{.}  \label{y1c}
\end{equation}

Thus%
\begin{equation}
1\ll y_{c}\ll \tilde{X}_{2}\text{.}  \label{1y}
\end{equation}

Then, the formula (\ref{gay}) gives us that 
\begin{equation}
\gamma \approx \frac{E_{1}-\sqrt{E_{1}^{2}-1}}{y_{c}}\tilde{X}_{2}\sim
x_{c}^{-1}  \label{yxc}
\end{equation}%
becomes unbounded in perfect agreement with (\ref{gac}). This is because $%
\tilde{X}_{2}$ grows with $x_{c}$ much faster than $y_{c}$. Thus the fact
that the quantity $X_{2}$ is not invariant under transformation (in contrast
to $X_{1}$) to another frame, plays a key role. More precisely, $X_{2}$ is
finite in the (\ref{br}) frame but becomes unbound in the (\ref{y}) frame.

It follows from (\ref{y}) that for $y\rightarrow \infty $ the proper
distance $l=\int \frac{dy}{\sqrt{1+y^{2}}}$ also diverges. By itself, this
is not exceptional. It is worth reminding that for extremal horizons the
proper distance to any point diverges, so when the BSW effect occurs near
such a horizon, the proper distance becomes unbound. In the frame (\ref{y}),
there is no horizon but this property persists.

To summarize the results of this section, there are two alternative
pictures. Either we have a metric with the horizon and two particles having
finite energies in the corresponding frames, one of particles being cortisol
or a metric without a horizon but one of particles has inbound energy.

\section{Kinematic properties}

There exists simple explanation of the BSW effect in the original frame like
(\ref{br}). Namely, for any initial conditions of particles' motion, their
relative velocity in the point of collision tends to that of light, so the
Lorentzian factor of relative motion $\gamma $ diverges. It is convenient to
show this using so-called zero angular momentum observers (ZAMO) \cite{72}.
Then\ (see eq. (29) of \cite{k}),%
\begin{equation}
X=E-q\varphi =\frac{N}{\sqrt{1-V^{2}}}\text{.}  \label{za}
\end{equation}

In the horizon limit $N\rightarrow 0$, $X$ remains separated from zero for a
usual particle, so it immediately follows from (\ref{za}) that $V\rightarrow
1$. For the critical one, $X\sim N$, so the factor $\frac{1}{\sqrt{1-V^{2}}}$
remains bounded, $V\neq 1$. Then, according to formulas of relativistic
transformation of velocities, the relative velocity $V_{rel}.\rightarrow 1$,
so $\gamma \rightarrow \infty $.

This explanation applies directly to the metric in the form (\ref{br}).
However, for the form (\ref{y}) it is not so obvious since $N\neq 0$. For
the critical particle, eq. (\ref{za}) with (\ref{xycr}) taken into account,
reads now%
\begin{equation}
E_{1}y=\frac{\sqrt{1+y^{2}}}{\sqrt{1-V_{1}^{2}}}\text{.}
\end{equation}

For any finite $E_{1}$, $V_{1}$ also remains finite. However, now we already
know from the previous section that, when one approaches the horizon, $%
y\rightarrow \infty $. Then, eq. (\ref{za}) turns into%
\begin{equation}
V_{1}=\sqrt{1-\frac{1}{E_{1}^{2}}}<1\text{,}  \label{v}
\end{equation}

where it is assumed that $E_{1}>1$.

For a usual neutral particle, in (\ref{za}) one should put $q_{2}=0$ and
substitute $\tilde{X}_{2}$ from (\ref{Y}). When we choose the point of
collision close to the horizon, $\frac{X_{2}}{y_{c}}\sim
x_{c}^{-1}\rightarrow \infty $ according to (\ref{p1}), (\ref{y1c}).
Therefore, it follows from (\ref{za}) that $V_{2}\rightarrow 1$. Thus the
critical and usual particles retain their properties in that in both frame $%
V_{2}\rightarrow 1$ for a usual particle and $V_{2}<1$ for the critical one.
Correspondingly, near the horizon the relative velocity of both particles
turns out to be close to 1 automatically.

Thus according to (\ref{xycr}), the critical particle has the finite energy
equal to $q$ in both frames. Meanwhile, if we want collision to occur near
the horizon, the trajectory of a suitable usual particle with a finite
energy in the frame (\ref{br}) maps to the trajectory with large energy in
the frame (\ref{t}). In both frames a usual particle approaches the horizon
with almost a speed of light but the reasons in both cases are different. In
the first case, it is consequence of equations of motion with a finite
energy, the velocity takes arbitrary values far from the horizon. In the
second one, the velocity is close to 1 for any finite $y$ due to a large
energy. We would like to stress that large energies in the frame (\ref{y})
arise not due to some additional assumption but simply due to general
properties of the metric plus requirements of collision between particles of
different kinds near the horizon.

The difference in properties of both particles clarify also, why the crucial
role in resolving the paradox was played by the behavior of the constant of
integration $t_{0}$. Now, this can be understood as follows. To make
collision possible, particles must meet in some point near the horizon. Let
us imagine that both particles are sent from one point at different moments
of time. Particle 1 (slow) should travel towards the horizon first and wait
there until particle 2 (rapid) starts its motion with some delay. As the
difference between velocities near the horizon is significant, the time lag
between moments of start (hence, $t_{0}$) should be also big. This explains
why $t_{0}$ becomes large when the point of collision approaches the horizon.

\section{Simplified example: Minkowski and Rindler metrics}

The essence of matter in the context under discussion can be also explained
if we resort to the simplest example - the flat space-time. By itself, this
case is trivial but it is a convenient tool to illustrate some subtleties
considered above. Let we have the metric%
\begin{equation}
ds^{2}=-dt^{2}x^{2}+dx^{2}\text{,}  \label{rin}
\end{equation}%
where we omitted the angular part. This is nothing than the Rindler metric.
For simplicity, we use the same letters $x,t$ as before but now, instead of (%
\ref{br}), $x$ (possibly, up to the sign) has the meaning of the proper
distance. We assume that $t\geq 0$ and $x\geq 0$, so we consider only one
quadrant of this space-time. One can introduce new coordinates \ $\tilde{t}$%
, $\tilde{y}$ in which the metric takes the most simple form - the Minkowski
one:%
\begin{equation}
ds^{2}=-d\tilde{t}^{2}+dy^{2}\text{.}  \label{min}
\end{equation}

Here,%
\begin{equation}
y=x\cosh t\text{,}  \label{ycos}
\end{equation}%
\begin{equation}
\tilde{t}=x\sinh t
\end{equation}%
and%
\begin{equation}
x^{2}=y^{2}-\tilde{t}^{2}\text{,}  \label{xyt}
\end{equation}%
\begin{equation}
\tanh t=\frac{\tilde{t}}{y}\text{.}
\end{equation}

It is seen from (\ref{xyt}) that the horizon $x=0$ corresponds to $y=\pm 
\tilde{t}$.

The Killing vector $\tilde{\xi}^{\mu }=(1,0)$ in the frame (\ref{min}).
Another Killing vector reads $\xi ^{\mu }=(1,0)$ in the Rindler coordinates (%
\ref{rin}). In the Minkowski frame (\ref{min}),%
\begin{equation}
\xi ^{\mu }=(y,\tilde{t})\text{.}  \label{xtil}
\end{equation}

Let us consider motion of geodesic particles (for simplicity, the mass of
each particle $m=1$). In the metric (\ref{min}), 
\begin{equation}
y=V\tilde{t}+y_{0},  \label{tv}
\end{equation}%
$V$ has the meaning of velocity, $t=\tau \tilde{E}$, $\tau $ is the proper
time,%
\begin{equation}
\tilde{E}=\frac{1}{\sqrt{1-V^{2}}}  \label{ev}
\end{equation}%
is the energy.

It follows from (\ref{xtil}) and (\ref{tv}) that the energies $E=-u_{\mu
}\xi ^{\mu }$ and $\tilde{E}=-u_{\mu }\tilde{\xi}^{\mu }$ are related
according to%
\begin{equation}
E=\tilde{E}y_{0}\text{.}  \label{ee}
\end{equation}

In our simplified model, the quantity (\ref{X}) $X=E$ for any particle. In
the Rindler coordinates, the trajectory (\ref{tv}) is rendered as 
\begin{equation}
x(\cosh t-V\sinh t)=y_{0}\equiv \frac{E}{\tilde{E}}\text{.}  \label{xx}
\end{equation}%
Using (\ref{ycos}) and (\ref{xx}), one finds%
\begin{equation}
y=\frac{E\tilde{E}(\alpha +V\sqrt{\alpha ^{2}-1})}{\alpha }\text{, }\alpha =%
\frac{E}{x}\text{. }  \label{ya}
\end{equation}

Let two particles collide in the point $x=x_{0}$. According to the general
formula (\ref{xz}), where now $q=0=\varphi $, the Lorentz factor of relative
motion%
\begin{equation}
\gamma =\frac{X_{1}X_{2}-Z_{1}Z_{2}}{x^{2}}\text{,}  \label{gaz}
\end{equation}%
where $Z_{i}=\sqrt{E_{i}^{2}-x^{2}}$.

As both energies are positive, $X_{i}\neq 0$ (i=1,2), and there are no
critical particles in our sense. However, if $X_{1}$ is small, let us call
the particle near-critical. Let collision occur near the horizon $x=0$, so $%
x_{0}$ is a small parameter. If for particle 1 (we call it near-critical)
the energy $E_{1}=X_{1}=O(x_{0})$ and the energy $E_{2}=O(1)$, it follows
from (\ref{gaz}) that $\gamma \sim x_{0}^{-1}$ grows unbounded when $%
x_{0}\rightarrow 0$.

From another hand, eq. (\ref{xz}) gives in the frame (\ref{min})%
\begin{equation}
\gamma =\tilde{E}_{1}\tilde{E}_{2}-\sqrt{\tilde{E}_{1}^{2}-1}\sqrt{\tilde{E}%
_{2}^{2}-1}\text{.}  \label{ge}
\end{equation}

There is no small denominator here, so one wonders how the BSW effect can be
explained.

Let, for simplicity, particle 2 have $V_{2}=0$, so in (\ref{ya}) $%
y_{2}=E_{2}=const$. It is seen from (\ref{ev}) that $\tilde{E}_{2}=1$, so (%
\ref{ge}) simplifies to $\gamma =\tilde{E}_{1}$.

We assume that collision occur at $x=x_{0}\ll 1$. Then, further information
follows from eq. (\ref{c2}) in which the expression for $y$ is taken from (%
\ref{ya}). Here, there are two typical cases.

1) Both particles are usual, $E_{1}\sim 1$, hence $\alpha \gg 1$. Then, $%
V_{1}<1$ is separated from zero, so $\tilde{E}_{1}=O(1)$, $\gamma \sim 1$,
there is no BSW effect.

2) $E_{1}\sim x_{0}\ll 1$, so $\alpha \sim 1.$ Then, $\tilde{E}_{1}\approx 
\frac{\tilde{E}_{2}}{E_{1}}\frac{\alpha }{\alpha +\sqrt{\alpha ^{2}-1}}\sim
x_{0}^{-1}$ grows unbound, and $\gamma $ does so. We obtain the BSW effect.

It is worth noting that in the frame (\ref{min}) it is the near-critical
particle that reaches the horizon with the speed approximately equal to that
of light, in contrast to the BR case (\ref{v}).

Obviously, this simplified model does not capture all features of the BSW
effect in the BR space-time since the horizon is nonextremal, a particle can
be only near-critical (not critical), etc. However, it illustrates the key
point: there exists a rather close analogy between the BSW effect from the
point of view of the Rindler observer in the flat space-time and in the BR
space-time. In both cases, the explanation can be suggested in terms of
motion of particles in the Kruskal-like / Minkowski frame. The BSW effect is
explained by the fact that a fast particle hits a slow (or motionless) one.
In turn, as the vicinity of the black hole extremal horizon can be
approximated by the BR metric, the very simple model of this section sheds
light on the essence of the BSW effect near black holes.

\section{Kruskal transformations and acceleration horizons verus black holes}

The fact that acceleration horizons can be used as a good approximation to
the metric near black holes, is important for understanding the nature of
the BSW effect. To elucidate relationship between two objects in the context
under discussion, let us exploit the Kruskal-type transformation. It is a
standard tool to pass from coordinates which are ill-defined on the black
hole horizon to the ones which are well-defined. Now, we apply such
transformation to the acceleration horizons. As this kind of transformation
is suited to black holes, this enables us to elucidate closed similarity of
the BSW effect in both types of the metrics in spite of their crucial
difference in the global structure of space-time. Let us rewrite the metric (%
\ref{br}) with $r_{+}=1$ in the form%
\begin{equation}
ds^{2}=x^{2}(-dt^{2}+dx^{\ast 2})+d\Omega ^{2}\text{,}  \label{a}
\end{equation}%
where 
\begin{equation}
x^{\ast }=-\frac{1}{x}.  \label{xk}
\end{equation}%
Introducing further the coordinates 
\begin{equation}
u=t-x^{\ast }\text{, }v=t+x^{\ast }  \label{uv}
\end{equation}%
and%
\begin{equation}
u=-\frac{2}{U}\text{, }v=-\frac{2}{V}\text{,}  \label{U}
\end{equation}%
we arrive at the metric%
\begin{equation}
ds^{2}=-4\frac{dUdV}{(V-U)^{2}}=-\frac{dT^{2}-dY^{2}}{Y^{2}}  \label{UV}
\end{equation}%
which is regular at the horizon (on the future horizon $U=0$, on the past
horizon $V=0$)$.$ Here, $U=T-Y$ and $V=T+Y$ similarly to (\ref{uv}), whence%
\begin{equation}
Y=-\frac{2}{x}(t^{2}-\frac{1}{x^{2}})^{-1}\text{, }T=-2t(t^{2}-\frac{1}{x^{2}%
})^{-1}.  \label{yy}
\end{equation}

The equations of motion in new coordinates read%
\begin{equation}
\dot{T}=\hat{X}Y^{2}\text{,}
\end{equation}%
\begin{equation}
\dot{Y}=-Y\sqrt{X^{2}Y^{2}-1}
\end{equation}%
for each particle. (When a particle moves towards $x=0$, the coordinate $Y<0$
and is increasing, so that $\dot{Y}>0$.)

Then, one can calculate the quantity (\ref{X}) in new coordinates $Y$, $T$
which we denote $\hat{X}.$ In this Section, the mass of any particle $m=1$.
As $\hat{X}$ $=u_{T}$, the standard rules for the transformation of the
components of the four-vector from the old frame to the new one give us the
following results.

For critical particle 1, eq. (\ref{cx}) holds in the old frame. Then, it
follows from (\ref{crit}), (\ref{tc}) and (\ref{yy}) that 
\begin{equation}
Y_{1}=-2x_{1}\lambda ^{2}<0  \label{yx1}
\end{equation}%
for the region under consideration where $x>0$. Also, direct calculations
gives us in the new frame%
\begin{equation}
\hat{X}_{1}=-\frac{E_{1}}{Y}>0\text{.}
\end{equation}%
For usual particle 2 with $q_{2}=0$,%
\begin{equation}
\hat{X}_{2}=\hat{E}_{2}=\frac{1}{2}(E_{2}^{-1}+E_{2}t_{0}^{2})\text{,}
\label{x2}
\end{equation}%
\begin{equation}
Y_{2}=\frac{-2x}{\left( xt_{0}+\sqrt{1-\frac{x^{2}}{E_{2}^{2}}}\right) ^{2}-1%
}\text{,}
\end{equation}%
where we used eqs. (\ref{xuE}), (\ref{tu}) in the old coordinate frame (\ref%
{br}). It is worth noting that $\hat{E}_{2}$ differs from (\ref{Y}) by an
unessential constant only. Then, direct calculations of the Lorentz factor
of relative motion (\ref{gamma}) gives us%
\begin{equation}
\gamma =E_{1}\hat{X}_{2}\left\vert Y\right\vert -\sqrt{E_{1}^{2}-1}\sqrt{%
\hat{X}_{2}^{2}Y^{2}-1}\text{.}  \label{gey}
\end{equation}

The condition of collision $Y_{1}=Y_{2}\equiv Y_{c}$ entails near the
horizon $x=0$ just eq. (\ref{t0}). If we want collision to occur at small $%
x_{c}$, the quantity $\left\vert Y_{c}\right\vert \sim x_{c}\sim t_{0}^{-1}$
is also small, $t_{0}$ grows unbound in accordance with (\ref{tx}). It
follows from (\ref{yx1}), (\ref{x2}) that $\hat{X}_{2}\left\vert
Y_{c}\right\vert \sim t_{0}.$ Then, eq. (\ref{gey}) gives us $\gamma \sim
t_{0}\sim x_{c}^{-1}$ in full agreement with (\ref{yxc}).

Let us now, instead of an acceleration horizon, have a true black hole
described by the metric (\ref{mf}). Following the standard route, we
transform it to%
\begin{equation}
ds^{2}=-fdudv+r^{2}d\Omega ^{2}\text{,}  \label{fu}
\end{equation}%
\begin{equation}
u=t-r^{\ast }\text{, }v=t+r^{\ast }\text{,}
\end{equation}%
where the tortoise coordinate%
\begin{equation}
r^{\ast }=\int \frac{dr}{f}.
\end{equation}

Further, one can introduce%
\begin{equation}
u=-\psi (-U)\text{, }v=\psi (V).
\end{equation}%
Then,%
\begin{equation}
ds^{2}=-gdUdV+r^{2}d\Omega ^{2}\text{, }g=f\psi ^{\prime }(-U)\psi ^{\prime
}(V)\text{,}  \label{g}
\end{equation}%
where we require that $g$ be finite on the horizon.

If the horizon extremal,%
\begin{equation}
f=x^{2}-ax^{3}+...
\end{equation}%
where $x=r-r_{+}$, $\ a$ is some constant and we assumed for simplicity that 
$f^{\prime \prime }(0)=2$, $r_{+}=1$. Then, 
\begin{equation}
r^{\ast }=-\frac{1}{x}(1+\xi )+O(1)\text{, }\xi =ax\ln x\text{.}  \label{xb}
\end{equation}%
The appearance of the term with $\xi $ is in agreement with eqs. 2.8, 2.9 of
Ref. \cite{lib}, where Kruskal-like coordinates were constructed explicitly
for the Reissner-Nordstr\"{o}m metric. It is negligible near the horizon, so
for small $x$ eq. (\ref{xk}) is a good approximation to (\ref{xb}). It is
sufficient to take $\psi (z)\approx -\frac{2}{z}$ near $x=0$, where $z$ is
the argument of the function. As compared to the acceleration horizon case,
now there are corrections due to $\xi $ and inconstancy of the coefficient $%
r^{2}$ at $d\Omega ^{2}$ in (\ref{fu}). If these small corrections are
discarded, the previous consideration of this Section applies. Then, in the
main approximation, we obtain the same formulas for the BSW effect.

Globally, the causal structures of the metrics (\ref{br}) and (\ref{mf}) are
very different. In particular, the acceleration horizon is completely
kinematic effect that appears or disappears depending on an observer, in
contrast to a true black hole. However, locally, both metrics are
indistinguishable in the immediate vicinity of the horizon in the main
approximation, the difference appears only due to small corrections away
from the horizon. Moreover, a Kruskal observer himself does not feel the
presence of the horizon locally unless he is making very subtle geometrical
experiments, so in this respect also there is no crucial difference between
both types of horizon. In a sense, the relationship between the frames (\ref%
{br}) and (\ref{y}) or (\ref{br}) and (\ref{UV}) is similar to the
relationship between (\ref{mf}) and (\ref{g}). Thus the entire picture is
self-consistent: as the BSW effect exists near black hole horizons, it also
exists for space-times whose metric is a good approximation to the black
hole metric near the horizon.

Meanwhile, the BSW effect arises just to collisions in a vicinity of the
horizon. Therefore, approximating the black hole metric by that of the
acceleration horizon, we obtain new explanation of the BSW effect near black
holes. It is given from the viewpoint of an observer who crosses the horizon
and is complimentary to the previous explanation \cite{k} which was done in
the frame of a static or stationary observer who is sitting in a fixed point
or is orbiting a black hole.

It is worth paying attention to the following circumstance. In the frame
where the horizon is absent or does not manifest itself explicitly, the
energy of a usual particle is unbound. Had we started from this frame, the
BSW effect would have looked trivial because of large $\hat{E}_{2}$ or $%
\tilde{E}_{2}$. However, the nontrivial fact is that in the original frames (%
\ref{mf}) or (\ref{br}), both energies $E_{1}$ and $E_{2}$ are finite. Then,
"trivializing" the BSW effect by passing in a new frame and revealing the
behavior of $\hat{E}_{2}$ can be considered as an explanation of the effect
which was not obvious in the original frame.

\section{Axially symmetric rotating metrics}

Many properties of the axially symmetric rotating acceleration horizons are
similar to those of the BR space-time.

Let us consider the metric%
\begin{equation}
ds^{2}=-dt^{2}N^{2}+g_{\phi \phi }(d\phi -\omega
dt)^{2}+g_{rr}dr^{2}+g_{\theta \theta }d\theta ^{2},  \label{m}
\end{equation}%
where the coefficients do not depend on $t$ and $\phi $. It is convenient
for further purposes to write $g_{rr}=\frac{C(\theta ,r)}{r^{2}}$. The Kerr
metric belongs just to this class. We choose the coordinate $r$ in such a
way that $r=0$ corresponds to the horizon, so for the extremal case $%
N^{2}\sim r^{2}$ for small $r$ by definition.

Then, 
\begin{equation}
N=A(\theta ,r)r  \label{na}
\end{equation}%
\begin{equation}
\omega =\omega _{H}+\bar{\omega}\text{, }  \label{om}
\end{equation}%
\begin{equation}
\bar{\omega}=-B(\theta ,r)r  \label{om1}
\end{equation}%
where $A(\theta ,r)$ and $B(\theta ,r)$ are regular in the vicinity of the
horizon, $A(\theta ,0)$, $B(\theta ,0)\neq 0$. The sign "minus" is chosen in
(\ref{om1}) since, say, for the Kerr metric $B(\theta ,0)>0$. The
presentation of the coefficient $\omega $ (\ref{om1}) follows from the fact
that for regular extremal black holes the first correction to $\omega _{H}$
near the horizon must have the order $N$ \cite{tan} which, in turn, is
proportional to $r$ according to (\ref{na}).

Then, the metric can be rewritten as%
\begin{equation}
ds^{2}=-dt^{2}N^{2}+g_{\phi \phi }(d\bar{\phi}-\bar{\omega}%
dt)^{2}+g_{ab}dx^{a}dx^{b},  \label{m2}
\end{equation}%
\begin{equation}
\bar{\phi}=\phi -\omega _{H}t\text{,}  \label{bar}
\end{equation}%
where $a,b=r,\theta $. The variable $\bar{\phi}$ corresponds to the frame
corotating with the horizon. Hereafter, we use the bar sign to denote
quantities in this frame. Near the horizon, we may describe the geometry
approximately, truncating the metric near the horizon and replacing the
coefficients $A,B,C$, $g_{\phi \phi }$ and $g_{\theta \theta }$ by their
limiting values at $r=0$. Then, we obtain%
\begin{equation}
ds^{2}=-A^{2}(\theta )r^{2}dt^{2}+g_{\phi \phi }(\theta )(d\bar{\phi}%
+B(\theta )rdt)^{2}+C(\theta )\frac{dr^{2}}{r^{2}}+g_{\theta \theta }(\theta
)d\theta ^{2}\text{.}  \label{ac}
\end{equation}%
where $A(\theta ,0)\equiv A(\theta )$, etc. The metric (\ref{ac}) belongs to
the class (\ref{m2}) with 
\begin{equation}
\bar{\omega}_{H}=0\text{.}  \label{oh}
\end{equation}

If we consider the metric (\ref{ac}) not as near-horizon approximation to (%
\ref{m}) or (\ref{m2}) that describes the black hole but, instead, as an
exact space-time, we obtain the metric of the acceleration horizon.
Alternatively, one can rescale the variables according to 
\begin{equation}
r=\varepsilon \tilde{r}\text{, }t=\frac{\tilde{t}}{\varepsilon }\text{,}
\label{var}
\end{equation}%
substitute them into (\ref{m}) or (\ref{m2}) and take the limit $\varepsilon
\rightarrow 0$. Then, we again obtain (\ref{ac}) with $r$ and $t$ replaced
by $\tilde{r}$, $\tilde{t}$.

In a somewhat different form, such a limiting transition was performed in 
\cite{hm}, \cite{b}. The resulting geometry is the generalization of the $%
AdS_{2}$x$S_{2}$ one. In particular, nontrivial manifold can be obtained for
the vacuum case when the metric (\ref{m}) describes the Kerr black hole.

\section{Equations of motion for rotating case}

Let a test particle move in the background (\ref{m}). Then, due to the
independence of the metric coefficients of $\phi $ and $t$, there are two
Killing vectors and two integrals of motion - energy $E$ and angular
momentum $L$. Hereafter, we restrict ourselves to the motion in the
equatorial plane $\theta =\frac{\pi }{2}$. Then, 
\begin{equation}
m\frac{dt}{d\tau }=\frac{X}{N^{2}}\text{, }  \label{ttr}
\end{equation}%
\begin{equation}
m\frac{d\phi }{d\tau }=\frac{L}{g}+\frac{\omega X}{N^{2}},  \label{phir}
\end{equation}%
\begin{equation}
m\frac{\sqrt{C}}{r}\frac{dr}{d\tau }=\varepsilon \frac{Z}{N}\text{, }%
\varepsilon =\pm 1\text{,}  \label{u1}
\end{equation}%
\begin{equation}
X=E-\omega L\text{, }Z=\sqrt{X^{2}-N^{2}(m^{2}+\frac{L^{2}}{g})},\text{ }%
g\equiv g_{\phi \phi }\text{,}  \label{zu}
\end{equation}%
where the metric coefficients are taken at $\theta =\frac{\pi }{2}$. In (\ref%
{u1}), we assume the minus sign that corresponds to a particle moving
towards the horizon.

In a similar manner, for (\ref{ac}) we obtain (\ref{tt}) - (\ref{u1}) with $%
X,Z\,$\ and $\phi $ replaced with $\bar{X}$ and $\bar{\phi}$.

For the metric (\ref{ac}), $N=Ar.$ The quantity in the original frame and
the one corotating with a black hole are related according to%
\begin{equation}
\bar{X}=\bar{E}-\bar{\omega}L=X\text{, }  \label{xe}
\end{equation}

\begin{equation}
\bar{E}=E-\omega _{H}L\equiv X_{H},  \label{eh}
\end{equation}%
\begin{equation}
L=\bar{L}.  \label{l}
\end{equation}%
\begin{equation}
\bar{Z}=Z=\sqrt{\bar{X}^{2}-N^{2}(m^{2}+\frac{L^{2}}{g})}.  \label{Z}
\end{equation}

\section{Classification of particles}

As usual, we assume the forward in time condition $\frac{dt}{d\tau }>0$.
This entails that $X=\bar{X}\geq 0$. If $X>0$ everywhere, we call such a
particle usual. Meanwhile, the forward in time condition admits also

\begin{equation}
X_{H}=0  \label{xh}
\end{equation}%
since $N=0$ on the horizon. In such a case, we call the particle critical.
Hereafter, subscript "H" means that the corresponding quantity is taken on
the horizon. Division of particles into these two classes is crucial for the
BSW effect \cite{ban}, \cite{k}.

For the metric (\ref{ac}), taking into account (\ref{oh}), (\ref{xh}) one
obtains that%
\begin{equation}
\bar{E}=0  \label{e0}
\end{equation}%
for critical particles and 
\begin{equation}
\bar{E}>0  \label{us}
\end{equation}%
for usual ones. In terms of unbarred quantities the condition of criticality
takes a form \cite{k}%
\begin{equation}
E-\omega _{H}L=0\text{.}  \label{cr}
\end{equation}

Then, one obtains from (\ref{Z}), (\ref{ga}) that 
\begin{equation}
\bar{X}=\bar{E}+BLr  \label{xu}
\end{equation}%
for a usual particle.

For the critical particle,%
\begin{equation}
\bar{X}=BLr.  \label{xc}
\end{equation}

After substitution into (\ref{ttr}), (\ref{phir}), one obtains that%
\begin{equation}
m\frac{dt}{d\tau }=\frac{\bar{E}+BLr}{A^{2}r^{2}}\text{, }  \label{tr}
\end{equation}%
\begin{equation}
m\frac{d\phi }{d\tau }=L(\frac{1}{g}-\frac{B^{2}}{A^{2}}).  \label{an}
\end{equation}

\section{BSW effect}

In the rotating case and nonzero angular momentum, calculation of the
Lorentz factor gives us 
\begin{equation}
\gamma =\frac{1}{m_{1}m_{2}}(c-d)\text{, }c=\frac{X_{1}X_{2}-Z_{1}Z_{2}\text{%
, }}{N^{2}}\text{, }d=\frac{L_{1}L_{2}}{g}  \label{ga}
\end{equation}%
that generalizes slightly (\ref{xz}). The only potential case of interest is
when one particle is critical (say, particle 1) and particle 2 is usual
since only such combination can give the BSW effect \cite{ban}, \cite{k}.

Then, by substitution into (\ref{ga}), one obtains that if collision takes
place near the horizon, so $N\rightarrow 0$,

\begin{equation}
E_{c.m.}^{2}\approx 2\frac{\left( \bar{E}_{2}\right) _{H}}{N}\left( \frac{%
BL_{1}}{A}-\sqrt{\frac{B^{2}}{A^{2}}L_{1}^{2}-m_{1}^{2}-\frac{L_{1}^{2}}{%
g_{H}}}\right) .  \label{3}
\end{equation}%
Thus in the horizon limit $E_{c.m.}^{2}$ grows indefinitely, so the BSW
effect manifests itself. Eq. (\ref{3}) has meaning for the angular momenta $%
L_{1}^{2}\geq m_{1}^{2}(\frac{B^{2}}{A^{2}}-\frac{1}{g_{H}})^{-1}$ only.
Otherwise, a critical trajectory cannot be realized and the BSW effect is
absent.

\section{Example: vacuum metric}

The simplest and, at the same time, physically relevant example, can be done
if the extremal Kerr metric is used as a "seed" one (\ref{m}): 
\begin{equation}
ds^{2}=-dt^{2}(1-\frac{2au}{\rho ^{2}})-\frac{4a^{2}u\sin ^{2}\theta }{\rho
^{2}}d\phi dt+\frac{\rho ^{2}}{\Delta }du^{2}+\rho ^{2}d\theta
^{2}+(u^{2}+a^{2}+\frac{2ua^{3}\sin ^{2}\theta }{\rho ^{2}})\sin ^{2}\theta
d\phi ^{2}\text{.}  \label{k}
\end{equation}%
Here, $u$ is the Boyer-Lindquiste coordinate, $\rho ^{2}=u^{2}+a^{2}\cos
^{2}\theta $, $\Delta =(u-a)^{2}$, $a$ characterizes the angular momentum of
a black hole. Then, the corresponding coefficients entering the metric (\ref%
{ac}) are equal to 
\begin{equation}
A=\frac{1}{2a}\sqrt{1+\cos ^{2}\theta }\text{, }B=\frac{1}{2a^{2}}\text{, }%
C=a^{2}(1+\cos ^{2}\theta )\text{,}  \label{ab}
\end{equation}%
\begin{equation}
g=\frac{4a^{2}\sin ^{2}\theta }{1+\cos ^{2}\theta }\text{.}  \label{gtt}
\end{equation}

For equatorial motion $\theta =\frac{\pi }{2}$, $A=\frac{1}{2a}=aB$, $%
C=a^{2} $, $g=4a^{2}$.

Performing the limiting transition based on the approaches of \cite{hm} and 
\cite{b} described in Sec. II, we arrive at the metric (\ref{ac}) with $%
r=u-a $,%
\begin{equation}
ds^{2}=\frac{(1+\cos ^{2}\theta )}{2}(-dt^{2}\frac{r}{r_{0}^{2}}^{2}+\frac{%
r_{0}^{2}dr^{2}}{r^{2}}+r_{0}^{2}d\theta ^{2})+\frac{2r_{0}^{2}\sin
^{2}\theta }{1+\cos ^{2}\theta }(d\phi +\frac{rdt}{r_{0}^{2}})^{2}\text{,}
\label{kr}
\end{equation}%
where $r_{0}^{2}=2a^{2}.$ This is just the form listed in eq. 2.6 of \cite{b}%
. In what follows, we put $r_{0}=1$, so $a^{2}=\frac{1}{2}$.

Performing the coordinate transformations given by eqs. (\ref{xty}), (\ref%
{tty}) with $x$ replaced with $r$ and 
\begin{equation}
\phi =\tilde{\phi}+\ln \frac{\cos \tilde{t}+y\sin \tilde{t}}{1+\sqrt{1+y^{2}}%
\sin \tilde{t}}\text{,}
\end{equation}%
we arrive at the metric given in eq. 2.9 of \cite{b}:%
\begin{equation}
ds^{2}=\frac{(1+\cos ^{2}\theta )}{2}[-d\tilde{t}^{2}(1+y^{2})+\frac{dy^{2}}{%
1+y^{2}}+d\theta ^{2}]+\frac{2\sin ^{2}\theta }{1+\cos ^{2}\theta }(d\tilde{%
\phi}+yd\tilde{t})^{2}\text{.}  \label{yl}
\end{equation}

Let us consider motion in the equatorial plane, $\theta =\frac{\pi }{2}$.
Equations of motion for $\frac{d\tilde{t}}{d\tau }$ and $\dot{y}$ have
exactly the same form (\ref{tt}), (\ref{yt}) where now%
\begin{equation}
\tilde{X}=\tilde{E}+\tilde{L}y\text{,}
\end{equation}%
\begin{equation}
\tilde{Z}=\sqrt{\tilde{X}^{2}-(1+y^{2})(\frac{\tilde{L}^{2}}{g}+m^{2})}\text{%
,}  \label{zyy}
\end{equation}%
tilde refers to the coordinate frame (\ref{yl}), $\tilde{L}=L$. It is
convenient to consider motion of critical and usual particles separately, as
before.

\subsection{Critical particle}

Then, solving equations (\ref{u1}), (\ref{tr}) with $\bar{E}=0$ and (\ref{ab}%
), (\ref{gtt}) taken into account, one obtains%
\begin{equation}
r=r_{0}\exp (-\lambda \tau )\text{, }\lambda ^{2}=3L^{2}-2  \label{rcr}
\end{equation}%
which is similar to (\ref{crit}) but with another value of $\lambda ,$%
\begin{equation}
t=\frac{2L}{r_{0}\lambda }\exp (\lambda \tau )=\frac{2L}{\lambda r}.
\label{tca}
\end{equation}

Here, it is assumed that $L>0$ in accordance with the forward in time
condition and $L>\sqrt{\frac{2}{3}}$.

For the critical particle, one can show, using the formulas of coordinate
transformation (\ref{xty}), (\ref{tty}) that the state with $\bar{E}=0$ in
the frame (\ref{kr}) maps on the state with $\tilde{E}=0$ as well. Then, it
follows from (\ref{yx}) that%
\begin{equation}
y=\frac{1}{2}(r+\frac{D}{r})\text{, }
\end{equation}%
\begin{equation}
D=\frac{4L^{2}}{\lambda ^{2}}-1=\frac{L^{2}+2}{\lambda ^{2}}>0\text{.}
\end{equation}

As $y>0$ and $L>0$, $\frac{d\tilde{t}}{d\tau }$ has the same sigh as $Ly$,
so the forward in time condition $\frac{d\tilde{t}}{d\tau }>0$ is satisfied.

\subsection{Usual particles}

To make the issue clearer, we assume additionally $L=\tilde{L}=0$ that is
similar to the condition that a particle is uncharged for the spherically
symmetric space-time case. Then, it follows from (\ref{u1}) and (\ref{Z})
that%
\begin{equation}
\dot{r}^{2}=4E^{2}-2r^{2}\text{,}
\end{equation}%
\begin{equation}
r=-\sqrt{2}E\sin \sqrt{2}\tau \text{,}
\end{equation}%
\begin{equation}
t=-\frac{1}{\sqrt{2}E}\cot \sqrt{2}\tau +t_{0}=\frac{1}{r}\sqrt{1-\frac{r^{2}%
}{2E^{2}}}+t_{0}\text{.}  \label{tma}
\end{equation}

In the point of collision $r_{c}$, times (\ref{tca}) and (\ref{tma}) should
coincide, whence for small $r_{c}$ 
\begin{equation}
t_{0}=\frac{1}{r_{c}}(\frac{2L}{\lambda }-1)+O(r_{c})>0\text{.}
\end{equation}

Eq. (\ref{yx}) gives us the formula 
\begin{equation}
y=-p\sin \sqrt{2}\tau +t_{0}\cos \sqrt{2}\tau
\end{equation}%
with 
\begin{equation}
p=\frac{1}{2}(E\sqrt{2}-\frac{1}{E\sqrt{2}}+t_{0}^{2}E\sqrt{2})
\end{equation}%
instead of (\ref{py}), (\ref{p}).

Thus the scheme developed for the BR space-time remains the same, only
numeric coefficients somewhat \ change. Correspondingly, the conclusions are
also the same. The main point is that in the near-horizon limit $t_{0}\sim
x_{c}^{-1}\sim y_{c}\rightarrow \infty $ and $\frac{\tilde{X}_{2}}{y_{c}}%
\sim t_{0}\rightarrow \infty $ as it was in the BR case. This leads to
divergences of $\gamma $ in this limit in the exactly the same manner as was
explained in Sec. IV. Thus for rotating acceleration horizon the BSW effect
is confirmed also in both frames.

\section{Comparison with previous statements in literature}

Recently, the paper \cite{g} appeared in which it is stated that the BSW
effect is absent in the metric of the acceleration horizon (instead, the
term "near-horzion extremal Kerr geometry (NHEK)" is used in \cite{g}). The
reasons of discrepancy between the main claims of \cite{g} and of our work
are quite simple. It was assumed in \cite{g} that energies of the colliding
particles $\bar{E}_{1},\bar{E}_{2}>0$. However, these particles, in our
terminology, are \textit{usual}. It was shown earlier \cite{k} in a more
general context, that collisions between two usual or two critical particles
cannot produce the BSW effect. The BSW effect requires one usual and one
critical particles. In doing so, the critical condition (\ref{e0}) reads $%
\bar{E}_{1}=0$ that does not fall into the class of collisions considered in 
\cite{g}, so the BSW effect was overlooked there.

The value $\bar{E}=0$ is rejected in the end of Sec. III of \cite{g} on the
basis of observation that with such a value of energy $\dot{r}$ grow
unbound. Indeed, it follows from (\ref{rcr}) \ that for $\tau \rightarrow
-\infty $ both $r$ and $\dot{r}$ tend to infinity, so the particle falls
towards the horizon from infinity. Although $\dot{r}$ becomes unbound, the
physical velocity \thinspace measured by a local observer remains finite and
less than the speed of light. This is connected with the fact that the
metric is not asymptotically flat. Indeed, consider for simplicity pure
radial motion in the spherically symmetric case. Then, $V=\frac{dl}{d\tau
_{loc.}}$, where $dl$ is the proper distance and $d\tau _{loc.}=\sqrt{f}dt$
is the proper time measured in the static frame. In the metric (\ref{mf}), $%
f=x^{2}$ for (\ref{br}) with $r_{+}=1$. After substitution of (\ref{crit}), (%
\ref{tc}), where it is implied that $E>1$, one obtains that $V=\sqrt{1-\frac{%
1}{E^{2}}}<1$ as it should be. (With minimum changes this applies also to
rotating metrics.) Thus there is nothing unphysical in such a behavior.
Moreover, in our context it is behavior near the horizon but not near
infinity which is relevant.

Although, on the first glance, condition (\ref{e0}) looks unusual, it is
equivalent to the standard condition of criticality (\ref{cr}) (so important
for the BSW effect \cite{ban}, \cite{k}) rewritten in the frame corotating
with the horizon. Moreover, the corresponding condition (\ref{xh}) is the
same for the original and rotating frames since $X$ is invariant under
transformation from one frame to another, so (\ref{e0}) and (\ref{cr}) are
different manifestation of the same property.

Also, it is stated in eq. (27) of \cite{g} that the near-horizon limit in
the original metric (\ref{k}) gives the critical relationship between the
parameters. Obviously, the integrals of motion do not depend on coordinates
since they are constants. Therefore, if the relation is not critical away
from the horizon, it cannot change it character and become critical near the
horizon. The origin of the mistake can be explained as follows. Let us
consider eq. (\ref{tr}) for the Kerr metric or its generalization (\ref{ac}%
). If one multiplies both sides by $r^{2}$ and takes the limit $r\rightarrow
0$, one naively "obtains" that $\bar{E}=0$. This is just the condition of
the criticality (\ref{e0}) or, equivalently, (\ref{cr}). However, it does
not mean that $\bar{E}>0$ is impossible. Instead, it only means that for $%
\bar{E}>0$ the quantity $\frac{dt}{d\tau }$ itself diverges, so $r^{2}\frac{%
dt}{d\tau }$ tends to $\frac{\bar{E}}{A^{2}}$ but not to zero. Had eq. (27)
of \cite{g} been universal, it would have meant that in the Kerr metric no
particles can move except from the critical ones.

Thus there is continuity between two kinds of the effect - due to black hole
and acceleration horizons. Let us stress once again that in the near-horizon
limit the geometry of any extremal black hole looks like an infinite throat
plus small unimportant corrections. It is impossible that with such small
deviations the quantity $E_{c.m.}$ be unbound near the horizon whereas in
the exact limit it would have suddenly become bounded. All this applies both
to the rotating and static cases, the latter being even simpler and clearer.

\section{Discussion and conclusions}

The results obtained in the present paper apply to two issues: (i) the
throat geometries as such when a black hole is absent but there is an
acceleration horizon, (ii) black holes since these throats serve as
near-horizon approximation. In case (ii) they give a new kinematic
explanation of the BSW effect itself - irrespective, whether it happens near
the black hole or acceleration horizon. In doing so, the frame (\ref{y}) in
which the metric is explicitly regular, is the direct analogue of the
Kruskal coordinate system and corresponds to an observer who can cross the
horizon. Then, we have two viewpoints complementary to each other based on
the original stationary system \cite{k} and its Kruskal-like counterpart
(discussed in the present paper). In one frame, there is a horizon at static
position $x=0$, the energies of particles are finite. If one particle is
critical, the other one is usual and an infinite growth of $E_{c.m.}$ occurs
that constitues the BSW effect. In the second frame, there is no such a
horizon but one of particles moves with the speed close to that of light, so
its energy is very large. Then, an infinite growth of $E_{c.m.}$ arises due
to collision of fast particle that hits a slow target. The features under
discussion were illustrated on the simplest example - collision of particles
in the flat space-time. \ Both viewpoints perfectly agree on the existence
of unbound $E_{c.m.}$ (hence, the BSW effect) but both disagree with \cite{g}%
.

We want to stress that it is the existence of the BSW effect in space-time
under discussion that makes the standard picture of the BSW effect \cite{ban}
self-consistent since (let us repeat it) any extremal black hole is
approximated by the metric of an acceleration horizon in the vicinity of the
true black hole horizon - i.e. just in the region which is responsible for
the BSW effect.

Present consideration revealed rather interesting moment. Although the BSW
effect looks as a local phenomenon in that $E_{c.m.}$ contains the
characteristics of trajectories (particles) in the point of collision only,
the preceding history comes to foreground when the system (\ref{y}) is
involved. The very fact that so different trajectories (critical and usual)
meet in the same point, impose severe restrictions on the constant $t_{0}$
that controls the interval between trajectories in a given point in space,
so it manifests itself as a kind of time nonlocality. In the original frame
it is "hidden" but its role becomes explicit when the energy in the frame (%
\ref{y}) is calculated. Usually, such things are skipped as unimportant but
in the present case, it is the value of this constant that affects the
energy and velocity of the particle in the (\ref{y}) system and is important
for the BSW effect.

Thus the full picture of the BSW effect includes not only dynamic properties
that give unbound $E_{c.m.}$ but also kinematic condition that makes
collision possible. Earlier, it was shown that for collision near the inner
black hole horizon infinite $E_{c.m.}$ is formally possible but kinematic
condition necessary for two particles to meet in the same point, prevents
collision with such $E_{c.m.}$ \cite{lake2} - \cite{inner}. Now, we saw that
even for the BSW effect near the event and/or acceleration horizon the
kinematic condition is also important.

In a sense, both local and nonlocal properties of a system manifest
themselves in the BSW effect. For the metric, it is the local properties
which ensure continuity between the BSW effect near black hole and
acceleration horizons. For particles, the collision by itself is a local
event, but some fine-tuning between trajectories of the critical and usual
particles is required to arrange this event just near the horizon (that is
indirect manifestation of nonlocality).

Bearing in mind the possible role of throats \cite{b} in the context of the
AdS/CFT and Kerr/CFT correspondences in quantum field theory, one can think
that the relevance of the BSW effect in such geometries can be of potential
interest not only in astrophysics but in particle physics as well.

\begin{acknowledgments}
I thank I. V. Tanatarov for interest to this work and useful comments.
\end{acknowledgments}

\section{Appendix: transformation of potential between two frames}

Here, we give explicit formulas for transformation of the potential between
frames (\ref{br}) and (\ref{y}). Let in the frame (\ref{br}) the potential
take the form (\ref{phi}). Using the standard formulas for transformation of
vectors, one obtains that after coordinate transformations (\ref{xty}), (\ref%
{tty}), the four-potential reads%
\begin{equation}
\tilde{\varphi}=-\frac{(\sqrt{1+y^{2}}+y\cos \tilde{t})\sqrt{1+y^{2}}}{\sqrt{%
1+y^{2}}\cos \tilde{t}+y}
\end{equation}%
\begin{equation}
\tilde{A}_{y}=\frac{\sin \tilde{t}}{\sqrt{1+y^{2}}\left( \sqrt{1+y^{2}}\cos 
\tilde{t}+y\right) }\text{.}
\end{equation}

It is convenient to make the gauge transformation $\tilde{A}_{\mu
}\rightarrow \tilde{A}_{\mu }-\partial _{\mu }f$, where%
\begin{equation}
\frac{\partial f}{\partial t}=-\frac{1}{\sqrt{1+y^{2}}\cos \tilde{t}+y},
\label{f1}
\end{equation}%
\begin{equation}
\frac{\partial f}{\partial y}=\frac{\sin \tilde{t}}{\sqrt{1+y^{2}}\left( 
\sqrt{1+y^{2}}\cos \tilde{t}+y\right) }.  \label{f2}
\end{equation}

Taking derivatives of (\ref{f1}) and (\ref{f2}) with respect to $y$ and $t$,
respectively, it is easy to check that eqs. (\ref{f1}) and (\ref{f2}) are
mutually consistent. Then, for the new potential $\tilde{A}_{y}=0$, $\tilde{%
\varphi}=-y$ that coincides with (\ref{fy}) up to the constant that can be
chosen at will.

\end{document}